\documentclass[11pt]{article}

\usepackage[preprint]{acl}

\usepackage{times}
\usepackage{latexsym}

\usepackage[T1]{fontenc}
\usepackage[utf8]{inputenc}

\usepackage{microtype}

\usepackage{inconsolata}

\usepackage{graphicx}

\title{Effective Reinforcement Learning for Agentic Search by Recycling Zero-Variance Queries During Training}

\author{ \bf
João Coelho$^{a,b}$, 
João Magalhães$^c$,   
Bruno Martins$^b$,  
Chenyan Xiong$^a$\\
$^a$ Language Technologies Institute, Carnegie Mellon University, United States \\
$^b$ Instituto Superior Técnico and INESC-ID, University of Lisbon, Portugal\\
$^c$ NOVA LINCS, NOVA School of Science and Technology, Portugal \\
\normalsize{jmcoelho@andrew.cmu.edu}
}

\usepackage{amsmath}
\usepackage{amssymb}
\usepackage{pifont}
\usepackage{booktabs}
\usepackage[table]{xcolor}   % \rowcolor (the [table] option is required)
\usepackage{arydshln}        % \cdashline (dashed midrule)
\usepackage{enumitem}
\usepackage{tabularx}

\definecolor{midnightgreen}{rgb}{0.0, 0.29, 0.33}

\begin{document}
\maketitle
\begin{abstract}
The use of GRPO-style algorithms has become the standard strategy for training LLM search agents under outcome-only rewards. With these algorithms, a query contributes to parameter updates only when its rollout group mixes successes and failures; all-correct (too-easy) and all-incorrect (too-hard) groups are zero-variance and waste rollout cost. Existing approaches treat zero-variance as a static property and either discard or pre-filter such groups. 
We hypothesize and empirically validate that queries flip between zero-variance and signal-bearing states as the policy evolves during training. Building on this intuition, we propose query recycling, which returns zero-variance groups to a mutable pool for future resampling, so that the effective training distribution co-evolves with the policy.
With the proposed technique, a 1.7B parameter model trained on synthetic data can reach 66.0 average Pass@1 accross seven multi-hop QA benchmarks, matching or surpassing systems with up to 7B parameters trained on benchmark-derived supervision.
%Analysis of recycling patterns shows that %approximately 40\% of the recycled queries were initially too-easy, indicating that recoverable signal arrives comparably from both policy improvement and drift.
Analysis of recycling patterns shows that recycled queries supply roughly three quarters of the effective batch by the end of training, with contributions split between recovery from policy improvement and policy drift.\footnote{Code and data available at a \href{https://github.com/cxcscmu/agentic_search_query_recycling}
{public github repo.}} 
\end{abstract}

\vspace{-0.15cm}
\section{Introduction}
\vspace{-0.15cm}

\begin{figure}[t!]
  \centering
  \includegraphics[width=\columnwidth]{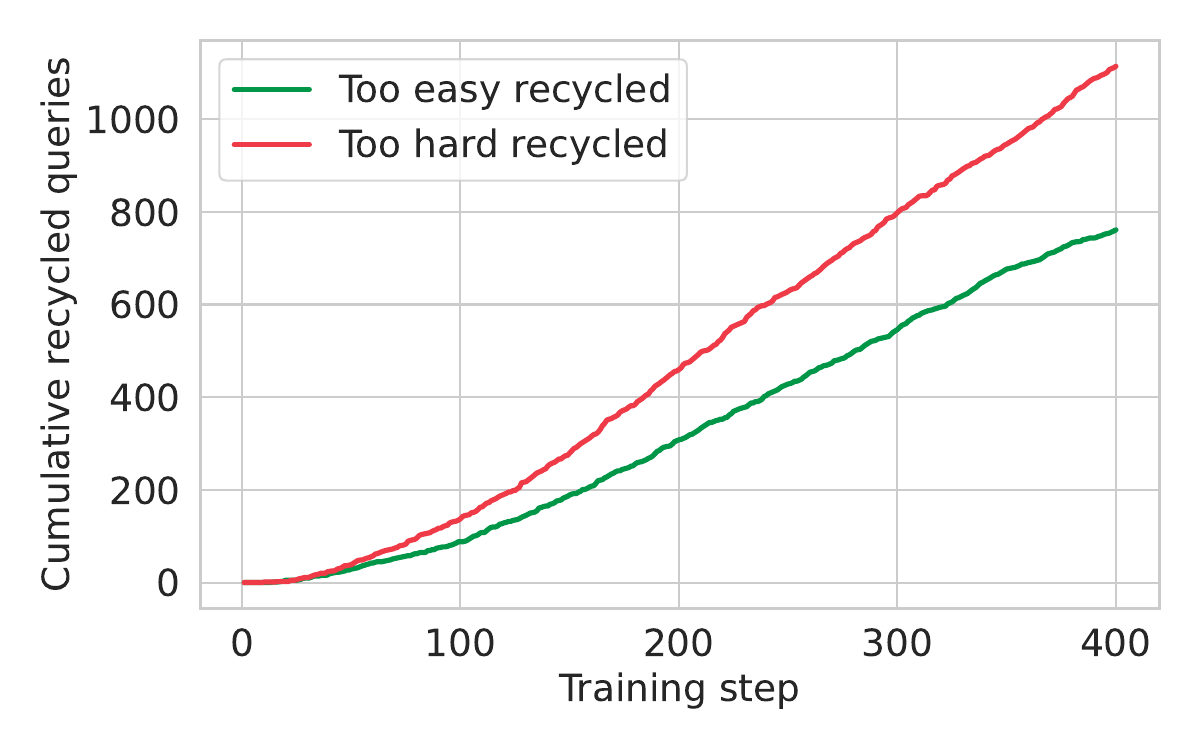}
  \vspace{-14pt}
  \caption{Recycling of zero-variance queries during RL training of a Qwen3-1.7B agent. A query is recycled if at some step $t$ its rollout group was too-easy (all correct) or too-hard (all incorrect), but when resampled at step $t' > t$ the same query was bearing signal.}
  \vspace{-10pt}
  \label{fig:rescue}
\end{figure}

Agentic search has emerged as a paradigm in which LLMs interleave reasoning with search calls to solve complex information-seeking tasks. Training LLMs as search agents increasingly relies on reinforcement learning~\cite{DBLP:journals/corr/abs-2505-15117}, with GRPO~\citep{shao2024grpo} or similar variants being the dominant algorithms. For each query in a training batch, the LLM policy produces a group of $K$ rollouts, and advantages are computed from the reward distribution within the group. Under outcome-only rewards, a query carries gradient only when its rollout group contains both successes and failures. Hence, queries the policy solves on every rollout (too-easy) or fails on every rollout (too-hard) produce no learning signal.

Prior work treats zero-variance as a static property. For instance, DAPO~\citep{DBLP:journals/corr/abs-2503-14476} oversamples candidate groups per step until the effective batch is full, and discards the zero-variance candidates. ASearcher~\citep{DBLP:journals/corr/abs-2508-07976} carries the same approach into search agent training. GRESO~\citep{DBLP:journals/corr/abs-2506-02177} goes further by using past epoch zero-variance statistics to probabilistically skip prompts before rollout, motivated by the observation (in the math reasoning domain) that the zero-variance status is temporally consistent across epochs. The inherent shared assumption across these methods is that a query's zero-variance status persists across training steps.

We instead hypothesize that zero-variance is a transient state, rather than a fixed property of the query. A query's zero-variance status reflects its success probability under the current policy, which shifts as the policy evolves. Thus, a query being zero-variance at training step $t$ should not entail the same status at step $t' > t$.

In the LLM search agent setting, we observe that queries flip in and out of zero-variance during training. Figure~\ref{fig:rescue} reports a Qwen3-1.7B run with recycling on a 10k-query pool, tracking queries that were zero-variance at some step $t$ and signal-bearing on resample at a later step $t' > t$. By the time of pool saturation, roughly 20\% of unique queries have flipped from zero-variance to signal-bearing. Queries re-enter the signal-bearing region from the too-hard tail as the policy improves, and from the too-easy tail as it drifts. Discarding zero-variance groups, or pre-filtering based on past statistics, eliminates this signal.

We formalize this intuition through dynamic pool management. We instantiate the training pool as a set of weighted queries from which we sample with probability proportional to the weights at each training step. The weights of sampled queries are then updated based on whether they produced gradient signal. Our approach, \textit{query recycling}, is the specific update rule that zeroes the weight of consumed signal-bearing queries while keeping zero-variance ones eligible for resampling. The same framework subsumes GRPO and DAPO as alternative weight-update rules over the same pool.

We conduct experiments in a fully synthetic and sandboxed training setup built on DeepResearchGym~\citep{2025_deep_research_gym}, without relying on the training splits of the considered evaluation benchmarks. We evaluate against standard GRPO and DAPO variants, training under matched rollout-budget regimes, and we find that recycling zero-variance queries is more effective than epoch-based replay of the training pool. Analysis further shows that recycled queries become a dominant source of effective training signal during late-stage GRPO, with roughly three-quarters of accepted groups originating from previously zero-variance regions of the pool by the end of training. Despite training exclusively on synthetic data, a Qwen3-1.7B agent reaches 66.0 average Pass@1 across seven multi-hop QA benchmarks, matching or surpassing prior systems of up to 7B parameters trained on benchmark-derived supervision.

% We will release all artifacts associated with this work, including training and query generation code, model checkpoints, and synthetic datasets.
   
\vspace{-0.15cm}
\section{Related Work}
\vspace{-0.15cm}

\paragraph{Search Agent Training.} Beyond test-time scaling, which allocates additional inference-time compute~\citep{snell2024scaling, muennighoff2025s1, anonymous2026diverse}, Search-R1~\citep{jin2025searchr1} established the now-standard paradigm of training LLM search agents with GRPO~\citep{shao2024grpo} under sparse outcome-only rewards, interleaving reasoning with multi-turn retrieval. Concurrent work followed the same recipe at various scales and model families~\citep{DBLP:journals/corr/abs-2505-15117, DBLP:journals/corr/abs-2503-05592, DBLP:conf/emnlp/ZhengFHCYLL25}, and outcome-based GRPO has since become the default starting point for search agent post-training.

Other authors argue that outcome-only rewards do not scale cleanly to long-horizon tool calling. As training progresses, agents exploit the sparse signal by issuing repetitive queries, since the objective provides no incentive for retrieval diversity or efficiency. SmartSearch~\citep{DBLP:journals/corr/abs-2601-04888} addresses this with a dual-level process reward that scores each query for novelty and usefulness. Fathom-DeepResearch~\citep{DBLP:journals/corr/abs-2509-24107} introduces a step-level reward that classifies each tool call by cognitive behavior and marginal utility.

BehaviorPrime~\citep{jin2026behaviorprime} follows prior work that uses SFT on trajectories from stronger teacher models to improve search agent performance~\citep{DBLP:conf/acl/ZengLLWLD024, DBLP:journals/corr/abs-2505-22648}. However, the authors filter the trajectories to instill good search behaviors. The paper further claims that this reduces the need for dense process rewards, which are otherwise prone to reward hacking.

% Other modifications target GRPO directly rather than its reward. Stratified GRPO~\citep{stratifiedgrpo, arxiv:2510.06214} shows that trajectories differing in the number and placement of search calls bias advantage estimation, and corrects for this by stratifying the baseline by call structure. Tree-GRPO~\citep{DBLP:journals/corr/abs-2509-21240} replaces chain rollouts with tree search over ReAct-style step nodes, achieving better exploration at lower rollout cost.

Training with live search APIs introduces cost and reproducibility concerns that a separate line of work tries to address, by replacing or sandboxing retrieval. ZeroSearch~\citep{DBLP:journals/corr/abs-2505-04588} fine-tunes an LLM to simulate a search engine. DeepResearchGym~\citep{2025_deep_research_gym} takes a different route, providing a static ClueWeb22 snapshot as a reproducible retrieval environment, and showing that agents trained on this environment generalize to live web search at inference. We build on DeepResearchGym and extend its synthetic query generation protocol.

\paragraph{Zero-Variance Mitigation Strategies.} Zero-variance groups were identified as a systematic inefficiency in GRPO training for mathematical reasoning~\citep{DBLP:journals/corr/abs-2503-14476}. When all rollouts for a query receive identical rewards, group-normalized advantages collapse to zero, yielding no policy-gradient signal despite incurring rollout cost. DAPO~\citep{DBLP:journals/corr/abs-2503-14476} addresses this problem through dynamic sampling, where data points with zero reward variance are discarded, and additional data points are sampled and rolled-out until the target number of informative samples is reached.

GRESO~\citep{DBLP:journals/corr/abs-2506-02177} predicts which prompts will be zero-variance using inter-epoch statistics and skips them before rollout, motivated by the observation that, in the math domain, zero-variance datapoints in one epoch tend to remain so in the next. Instead of skipping zero-variance groups, RL-ZVP~\citep{DBLP:journals/corr/abs-2509-21880} shapes their advantages, giving each rollout the group's outcome sign and a magnitude scaled by token entropy.

In the search-agent setting, ASearcher~\citep{DBLP:journals/corr/abs-2508-07976} applies DAPO-style post-rollout filtering, paying rollout cost and discarding zero-variance queries. DAVID-GRPO~\citep{DBLP:journals/corr/abs-2601-21699} targets all-failed zero-variance groups by finding the best trajectory in the group, truncating at the last step where valid evidence was retrieved, and resampling forward. This assumes that the current policy may solve the query but failed by chance.

\paragraph{Synthetic Data for Search Agents.} Recent work synthesizes training data for LLM search agents, often distilling trajectories from stronger reasoning models~\citep{sun2025simpledeepsearcher, DBLP:journals/corr/abs-2505-22648}. Several methods instead target structurally challenging tasks. For instance, WebSailor~\citep{li2025websailornavigatingsuperhumanreasoning} samples and fuzzifies entity graphs, while WebShaper~\citep{tao2025webshaperagenticallydatasynthesizing} composes complex QA samples through set-theoretic constructions. In turn, WebExplorer~\citep{liu2025webexplorerexploreevolvetraining} and ASearcher~\citep{DBLP:journals/corr/abs-2508-07976} obfuscate identifying details from factual questions to force longer search chains. WebResearcher~\citep{qiao2025webresearcherunleashingunboundedreasoning} escalates complexity through tool-augmentation, while ORBIT~\citep{thakur2026orbit} enforces answer verifiability through self and external checks.   
\vspace{-0.15cm}
\section{Methodology}
\vspace{-0.15cm}

In this section, we describe our training pipeline. We cover the query generation procedure (\S\ref{sec:query-generation}), the agent design and its tool interface (\S\ref{sec:agent-design}), the supervised fine-tuning stage that produces our RL starting point (\S\ref{sec:sft}), and the RL stage in which we introduce the query recycling strategy (\S\ref{sec:rl}).

\subsection{Query Generation}
\label{sec:query-generation}

% \paragraph{Pool Composition.}
% Our synthetic pool contains the 9k query-answer pairs from the DeepResearchGym synthetic dataset~\citep{2025_deep_research_gym}, which are grounded in ClueWeb22. We augment this initial set with 16k additional queries produced by the three methods described below, grounded in the same corpus. 

\paragraph{Pool Composition.} Our synthetic pool contains the 9k query-answer pairs from the DeepResearchGym synthetic dataset~\citep{2025_deep_research_gym}, which are grounded in ClueWeb22. We augment this initial set with 16k additional queries produced by three different methods, grounded in the same corpus. Each method aims to induce a different search strategy: webgraph queries probe hop-following along document adjacency, iterative-search queries train retrieval-driven chaining beyond the hyperlink structure, and comparative queries induce parallel retrieval over independently sourced entities.

\paragraph{Webgraph-based Queries.}
From a root document sampled uniformly from the corpus, we construct a subgraph by iteratively selecting either an in-link or an out-link of the current node, extending the random walk until reaching $n$ documents. The resulting documents are passed to a generator LLM in a single call, which produces one multi-hop question whose answer is a single entity, date, or number and that cannot be resolved from any single document. The hop structure is determined by hyperlink adjacency, and no retrieval is performed.

\paragraph{Iterative Search Queries.}
This method also constructs iterative chain questions, but replaces hyperlink traversal with dense retrieval, allowing it to reach documents that are not directly connected in the hyperlink graph. From a root document, the generator iteratively produces follow-up retrieval queries whose results are incorporated into progressively harder multi-hop questions, extending the chain until $n$ documents are reached.

\paragraph{Comparative Search Queries.}
The previous two methods construct sequential chains in which each hop depends on the previous one. Comparison questions instead require retrieving several entities independently before aggregating them, introducing a parallel retrieval structure. Starting from a root entity, the generator emits queries targeting entities that share a comparable attribute until $n$ entities are gathered, and constructs a question whose answer depends on reasoning over them.

\begin{table*}[t]
\centering
\small
\renewcommand{\arraystretch}{1.25}
\setlength{\tabcolsep}{8pt}
\begin{tabularx}{\textwidth}{@{}l X p{4cm}@{}}
\toprule
\textbf{Source} & \textbf{Query} & \textbf{Relevant Documents} \\
\midrule
DRGym &
Which three European football clubs did the Surinamese-born Dutch midfielder who later managed AC Milan and was honored as a Nelson Mandela Foundation Legacy Champion represent when he won the UEFA Champions League during his playing career? \newline
\textit{Answer:} Ajax, Real Madrid, AC Milan &
$-$ Clarence Seedorf \newline
$-$ Nelson Mandela \\
\addlinespace[3pt]
\cdashline{1-3}
\addlinespace[3pt]
Iterative &
What is the name of the game engine utilized by the open-source real-time strategy project to which the creator of Apache Spark contributed water rendering physics during his undergraduate studies? \newline
\textit{Answer:} Pyrogenesis &
$-$ Databricks \newline
$-$ Matei Zaharia \newline
$-$ 0 A.D. (video game) \\
\addlinespace[3pt]
\cdashline{1-3}
\addlinespace[3pt]
Webgraph &
The author of the 1991 Booker Prize-shortlisted novel \textit{Time's Arrow} also wrote a 1975 parody of country-house mysteries. What is the title of the novel for which that author's father won the Booker Prize in 1986? \newline
\textit{Answer:} The Old Devils &
$-$ Dead Babies (novel) \newline
$-$ Time's Arrow (novel) \newline
$-$ Kingsley Amis \\
\addlinespace[3pt]
\cdashline{1-3}
\addlinespace[3pt]
Comparative &
Among the King of Qin who became the first emperor of China, the Macedonian king who founded Alexandria, the Egyptian pharaoh whose nearly intact tomb was discovered by Howard Carter, and the last emperor of the Qing Dynasty who abdicated in 1912, who was the youngest at the time of their first accession to their respective thrones? \newline
\textit{Answer:} Puyi &
$-$ Alexander the Great \newline
$-$ King Tutankhamun \newline
$-$ Qin Shi Huang \newline
$-$ Puyi \\
\bottomrule
\end{tabularx}
\caption{Example queries from each generation source, shown together with gold answers and  supporting documents.}
\label{tab:query-examples}
\end{table*}

\paragraph{Query Filtering.}
All candidate queries pass an LLM-based validation stage that discards questions that are answerable from common knowledge or malformed. When constructing a question with $n$ target documents, we retain the intermediate version at every $n' < n$, so a single trace yields multiple valid questions of increasing complexity. Failed traces retain the last valid intermediate version, introducing a bias toward lower hop counts.

\begin{figure}[t]
  \centering
  \includegraphics[width=\columnwidth]{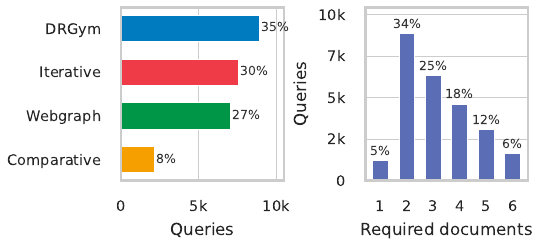}
  \caption{Distribution of the synthetic query pool according to the generation source (left) and the number of required supporting documents (right).}
  \label{fig:pool-distribution}
\end{figure}

Figure~\ref{fig:pool-distribution} shows the final distribution of questions by generation source and required document count. From the resulting pool, we fix two disjoint subsets used throughout all experiments: 15k QA pairs for supervised fine-tuning and 10k for reinforcement learning. Table~\ref{tab:query-examples} shows examples of queries generated by the aforementioned methods.

\subsection{Agent Architecture}
\label{sec:agent-design}

Our agent follows the ReAct-style design of AReal’s Tongyi DeepResearch implementation~\citep{DBLP:journals/corr/abs-2505-24298, DBLP:journals/corr/abs-2510-24701, DBLP:conf/iclr/YaoZYDSN023}, with two simplifications. First, we expose only a \texttt{search} tool, which allows the model to issue up to three parallel queries per turn. Second, we remove both (i) the URL-visiting tool and (ii) the auxiliary model used to extract information from webpage contents. Instead, the \texttt{search} tool directly returns cleaned text passages from relevant documents to the agent. The agent iteratively searches and reasons over the retrieved results until it produces a final answer or reaches a predefined step budget.

The \texttt{search} tool is backed by two retrieval sources depending on the setting. During training, queries are served by the DeepResearchGym API~\cite{2025_deep_research_gym, ning2026agenticsearchwildintents} over a ClueWeb22 snapshot~\cite{DBLP:conf/sigir/OverwijkXC22}, which gives a reproducible and sandboxed retrieval environment. During inference, we replace the backend with SERPER~\citep{serper2025api} live web search, allowing us to test whether policies trained in DeepResearchGym transfer to live web retrieval.

\subsection{Supervised Fine-Tuning}
\label{sec:sft}
Following prior work~\citep{jin2026behaviorprime}, we produce the RL starting checkpoint with a short Supervised Fine-Tuning (SFT) stage on trajectories distilled from a stronger teacher. For the 15k queries drawn from the synthetic pool described in \S\ref{sec:query-generation}, we run the agent framework of \S\ref{sec:agent-design} with a GPT-4.1-mini backbone as the teacher, and collect one trajectory per query. We retain 6.3k trajectories that follow the required action format and end with the correct answer for SFT. The same SFT checkpoint is used as the starting point for every configuration in our experiments, isolating the effect of the RL strategy.

\subsection{RL with Query Recycling}
\label{sec:rl}

\paragraph{Standard GRPO.} We train the SFT checkpoint with GRPO following the AReaL implementation~\citep{DBLP:journals/corr/abs-2505-24298}. Given a query $q$, the behavior policy produces a group of $K$ rollouts, $G_t(q)=\{o_1,\ldots,o_K\}$, where $t$ denotes the training step at which the group is sampled. Each rollout receives a binary outcome reward $r_i(q) = R(o_i,q) \in \{0,1\}$, indicating whether the final answer is correct.

We then compute group-normalized rollout advantages from the outcome rewards:
\[
A_i(q)
=
\frac{
r_i(q)-\mu_G
}{
\sigma_G+\epsilon_{\mathrm{norm}}
},
\qquad
\mu_G
=
\frac{1}{K}\sum_{j=1}^{K} r_j(q),
\]
where $\sigma_G$ is the standard deviation of the rewards in the group, and $\epsilon_{\mathrm{norm}}$ is a small numerical constant. The rollout-level advantage $A_i(q)$ is assigned to every valid token in rollout $o_i$. The policy is optimized with the following objective:
\begin{equation*}
\begin{aligned}
    \mathcal{J}(\theta) = \mathbb{E}_{q, G} \Bigg[ & \frac{1}{\sum_{i=1}^{K} L_i} \sum_{i=1}^{K} \sum_{\tau=1}^{L_i}
     \min\big( \rho_{i,\tau}\, A_i, \\
     & \quad \mathrm{clip}(\rho_{i,\tau}, 1-\epsilon, 1+\epsilon)\, A_i \big) \Bigg] \;,
\end{aligned}
\end{equation*}
where $L_i$ is the number of valid tokens in the rollout $o_i$, and where the per-token importance ratio $\rho_{i,\tau}$ between the current policy $\pi_{\theta}$ and the behavior policy $\pi_{\theta_{old}}$ is defined as follows:

\[
\rho_{i,\tau}
=
\frac{
\pi_\theta(o_{i,\tau}\mid q,o_{i,<\tau})
}{
\pi_{\theta_{\mathrm{old}}}(o_{i,\tau}\mid q,o_{i,<\tau})
} \;.
\] 

\paragraph{Dynamic Pool Management.}

Under the previous setting, a query $q$ contributes gradient signal only when its rollout group contains both successful and unsuccessful trajectories (i.e., $0 < \sum_{i=1}^{K} r_i(q) < K$). Let $p_t(q)$ denote the success probability of query $q$ at training step $t$:

\begin{equation}
p_t(q)
=
\mathbb{E}_{o \sim \pi_{\theta_t}(\cdot \mid q)}
\left[
R(o,q)
\right] \;.
\end{equation}
 
\noindent Assuming the $K$ rollouts in a group are conditionally independent given $q$, the probability that the group is zero-variance is given by:
\begin{equation} \label{eq:pzv}
P_{\mathrm{zv}}(q;\pi_{\theta_t}) = p_t(q)^K + (1-p_t(q))^K.
\end{equation}

Evaluating a query with a finite rollout group therefore yields a stochastic observation of whether the query is zero-variance under the current policy. Observing a zero-variance group at step $t$ does not imply that the query is intrinsically uninformative, nor that it will remain so under future policies. Since $P_{\mathrm{zv}}(q;\pi_{\theta_t})$ depends on the evolving policy, the zero-variance status is fundamentally non-stationary. A query observed as all-failed under an earlier policy may later become partially solvable as capabilities improve, while a previously all-correct query may re-enter the signal-bearing region as the policy drifts.

We exploit this non-stationarity through dynamic pool management. We formalize the training pool as a set of $N$ weighted queries:
\[
\mathcal{Q}^{(t)} = \{(q_i, w_i^{(t)})\}_{i=1}^{N}, \qquad w_i^{(0)} = 1,
\]
where $w_i^{(t)} \in [0,1]$ is the sampling probability of $q_i$ at step $t$. At each training step, we draw a candidate set $C_t$ of distinct $kB$ queries from $\mathcal{Q}^{(t)}$, with each query drawn with probability proportional to $w^{(t)}$, where $B$ is the training batch size and $k \geq 1$ is an oversampling factor. We then generate $K$ rollouts per candidate under $\pi_{\theta_t}$ and identify $\tilde{S}_t$, i.e. all signal-bearing queries:
\begin{equation}
\tilde{S}_t
=
\left\{
q \in C_t
\;\middle|\;
0 < \sum_{i=1}^{K} r_i(q) < K
\right\}.
\end{equation}

From $\tilde{S}_t$, we form the gradient update batch $S_t \subseteq \tilde{S}_t$ of size $\min(|\tilde{S}_t|, B)$, so that the effective batch size never exceeds $B$. Different training strategies correspond to different choices of the oversampling factor $k$ and different update rules for the query weights $w_i^{(t)}$.

\noindent\textbf{Query Recycling}. Our strategy zeroes the weight of the consumed signal-bearing queries $S_t$, effectively removing them from the pool: %Our strategy updates the weights of queries that contribute gradient signal to 0, effectively removing them from the pool:
\[
w_q^{(t+1)} =
\begin{cases}
0 & \text{if } q \in S_t \;, \\
1 & \text{otherwise} \;.
\end{cases}
\]

Queries that do not contribute gradient signal, including zero-variance groups and excess signal-bearing candidates not selected into $S_t$, remain eligible for sampling under future policies. Thus, the effective training distribution co-evolves with the policy $\pi_{\theta_t}$ rather than following a fixed multi-epoch traversal of the query pool.

Oversampling and recycling address complementary inefficiencies. Increasing $k$ raises the probability of observing signal-bearing queries within a training step, while recycling reallocates future sampling mass toward queries that have not yet contributed gradient updates. We use both mechanisms jointly throughout our experiments.

Existing methods can also be formalized within this framework, arriving from different choices of $k$ and weight updates. For example, GRPO fixes $k=1$ and consumes all sampled candidate queries, irrespective of gradient signal:
\[
w_q^{(t+1)} =
\begin{cases}
0 & \text{if } q \in C_t \;, \\
1 & \text{otherwise} \;.
\end{cases}
\]
When $\sum_{i=1}^{N} w_i^{(t)} = 0$, the pool is reset by setting $w_i \leftarrow 1$ for all queries, recovering standard multi-epoch iteration over a static dataset.

DAPO dynamically increases $k$ within a training step until $|\tilde{S}_t| = B$. All sampled candidates are then consumed using the same update rule as GRPO. This guarantees a full effective batch at the cost of unbounded rollout computation per step.

Bounded-DAPO fixes $k > 1$, avoiding the unbounded resampling loop of DAPO while allowing the effective batch size to vary when $|\tilde{S}_t| < B$. The weight update is otherwise identical to GRPO.

\vspace{-0.15cm}
\section{Experiments}
\vspace{-0.15cm}

This section evaluates the impact of query recycling. We describe the experimental setup (\S\ref{sec:exp_method}), report results across multiple QA benchmarks (\S\ref{sec:exp_results}), and analyze recycling dynamics (\S\ref{sec:analysis}).

\vspace{-0.15cm}
\subsection{Experimental Setup}
\label{sec:exp_method}
\vspace{-0.15cm}

We train Qwen3-1.7B and Qwen3-4B independently in two stages. First, we perform Supervised Fine-Tuning (SFT) on synthetic search trajectories to instill baseline retrieval behaviors. We then apply GRPO over the synthetic query pool described in Section~\ref{sec:query-generation}, constructed from ClueWeb22.

% We instantiate the procedure in Section~\ref{sec:rl} with batch size $B=32$, group size $K=4$, and oversampling factor $k \in \{1, 2\}$. Standard GRPO without recycling corresponds to $k=1$ in our setting. We evaluate two regimes: (Regime B) matched rollouts, which fixes the total number of trajectories to isolate per-rollout efficiency, and (Regime A) matched updates, which fixes the number of gradient steps, corresponding to the horizon at which the $k=1$ configuration exhausts the query pool once. %Configurations with $k>1$ without recycling exhaust the pool earlier under this setting.

We instantiate the procedure in Section~\ref{sec:rl} with batch size $B=32$, group size $K=4$, and oversampling factor $k \in \{1, 2\}$. Setting $k=1$ without recycling reduces to standard GRPO; $k=2$ without recycling corresponds to Bounded-DAPO; our full method uses $k=2$ with recycling. All configurations share a matched rollout budget of 84.5K trajectories, with $k=1$ run for 660 gradient steps and $k=2$ for 330. For the non-recycling baselines, this is equal to two epochs over the training pool.

We evaluate using the BehaviorPrime setup~\cite{jin2026behaviorprime}. For multi-hop question answering, we use 2WikiMultiHopQA~\citep{DBLP:conf/coling/HoNSA20}, Bamboogle~\citep{DBLP:conf/emnlp/PressZMSSL23}, HotpotQA~\citep{DBLP:conf/emnlp/Yang0ZBCSM18}, MuSiQue~\citep{DBLP:journals/tacl/TrivediBKS22}, Natural Questions~\citep{DBLP:journals/tacl/KwiatkowskiPRCP19}, PopQA~\citep{DBLP:conf/acl/MallenAZDKH23}, and TriviaQA~\citep{DBLP:conf/acl/JoshiCWZ17}. For web navigation, we consider GAIA~\citep{DBLP:conf/iclr/MialonF0LS24}, HLE~\citep{phan2025humanitysexam}, and WebWalkerQA~\citep{DBLP:conf/acl/0007Y0WXFZ0ZXH25}. We use GPT-4.1-mini as the judge for all benchmarks.

Inference uses Qwen3~\citep{DBLP:journals/corr/abs-2505-09388} default decoding parameters (temperature 0.6, top-$p$ 0.95, top-$k$ 20). We report Pass@1 averaged over eight runs. We compare against SearchR1, DeepResearcher, R1Searcher, BehaviorPrime, ASearcher-Web-7B, ORBIT, and SmartSearch. For the first four baselines, we report numbers from BehaviorPrime under matched evaluation, and for the others we reproduce results under comparable settings.

\vspace{-0.15cm}
\subsection{Empirical Results}\label{sec:exp_results}
\vspace{-0.15cm}

\begin{table*}[t!]
\centering
\small
\setlength{\tabcolsep}{4pt}
\renewcommand{\arraystretch}{1.15}
\begin{tabular}{lllcccccccccc}
\toprule
Model & Size & $k$ & Rollouts & 2Wiki & Bamb & HQA & MSQ & NQ & PopQA & TriQA & Avg \\
\midrule
\rowcolor{gray!15}
\multicolumn{12}{l}{\textbf{Prior Work} (training data includes MHQA benchmark splits)} \\
Search-R1       & 7B   & -- & --    & 47.9 & 57.6 & 63.0 & 27.5 & 60.0 & 47.0 & 76.2 & 54.2 \\
R1-Searcher     & 7B   & -- & --    & 65.8 & 65.6 & 53.1 & 25.6 & 52.3 & 43.4 & 79.1 & 55.0 \\
DeepResearcher  & 7B   & -- & --    & 66.6 & 72.8 & 64.3 & 29.3 & 61.9 & 52.7 & 85.0 & 61.8 \\
SmartSearch     & 3B   & -- & --    & \underline{78.7} & \underline{74.4} & 64.1 & 35.4 & 67.2 & 51.6 & 85.7 & 65.3 \\
BehaviorPrime   & 1.7B & -- & --    & 73.8 & 70.4 & 67.0 & 30.7 & \underline{73.6} & \textbf{56.6} & 86.9 & 65.6 \\
ORBIT           & 4B   & -- & --    & 75.2 & \underline{74.4} & 65.6 & 35.9 & 68.4 & 55.0 & 86.3 & 65.8 \\
ASearcher-Web   & 7B   & -- & --    & 76.4 & 71.2 & \underline{68.0} & 36.1 & 70.1 & 54.7 & 87.5 & \underline{66.3} \\
\midrule
\rowcolor{gray!15}
\multicolumn{12}{l}{\textbf{Ours} (full synthetic training over ClueWeb22 and evaluation using SERPER)} \\
\addlinespace[3pt]
\cdashline{1-12}
\addlinespace[3pt]
Qwen3                        & 1.7B & -- & --    & 37.4 & 44.4 & 44.1 & 16.3 & 54.5 & 41.8 & 68.6 & 43.9 $\pm$ 0.40 \\
\addlinespace[3pt]
\cdashline{1-12}
\addlinespace[3pt]
\quad + SFT                  & 1.7B & -- & --    & 54.8 & 47.2 & 54.4 & 24.6 & 67.6 & 52.0 & 86.0 & 55.2 $\pm$ 0.33 \\
\quad + GRPO                 & 1.7B & 1  & 84.5K & 70.9 & 68.8 & 66.6 & 36.9 & 69.9 & 54.4 & 86.9 & 64.9 $\pm$ 0.21 \\
\quad + B-DAPO               & 1.7B & 2  & 84.5K & 69.5 & 67.2 & 66.9 & 36.9 & 69.5 & 53.5 & 85.2 & 64.1 $\pm$ 0.20 \\
\quad + B-DAPO w/ recycling  & 1.7B & 2  & 84.5K & 71.7 & 72.4 & 67.7 & \underline{38.0} & 70.1 & 54.2 & \underline{87.8} & 66.0 $\pm$ 0.17 \\
%\multicolumn{12}{l}{\textit{Scaling up model size}} \\
\addlinespace[3pt]
\cdashline{1-12}
\addlinespace[3pt]
Qwen3                        & 4B   & -- & --    & 57.9 & 61.2 & 57.6 & 29.8 & 72.0 & 55.6 & 84.2 & 59.8 $\pm$ 0.24 \\
\addlinespace[3pt]
\cdashline{1-12}
\addlinespace[3pt]
\quad + SFT                  & 4B   & -- & --    & 68.9    & 64.4    & 64.5    & 33.8    & 73.1    & 55.7    & 88.5    & 64.1 $\pm$ 0.19  \\
\quad + GRPO                 & 4B   & 1  & 84.5K & 77.7 & 77.2 & \underline{73.0} & 40.7 & 69.9 & 53.1 & 89.1 & 68.7 $\pm$ 0.17 \\
\quad + B-DAPO               & 4B   & 2  & 84.5K & 77.4 & \underline{77.7} & 72.4 & \textbf{41.8} & \underline{73.7} & 54.2 & \textbf{90.8} & \underline{69.7} $\pm$ 0.17 \\
\quad + B-DAPO w/ recycling  & 4B   & 2  & 84.5K & \textbf{79.1} & \textbf{84.0} & \textbf{74.4} & \underline{41.3} & \textbf{74.4} & \underline{55.8} & \underline{89.9} & \textbf{71.3} $\pm$ 0.20 \\
\bottomrule
\end{tabular}
\vspace{-0.15cm}
\caption{Pass@1 on seven multi-hop QA benchmarks. All RL configurations start from the SFT checkpoint and use a matched rollout budget, equivalent to two epochs over the query pool for GRPO and DAPO variants. Averaged results over multiple runs are reported as mean $\pm$ standard error.}
\vspace{-0.15cm}
\label{tab:main-results}
\end{table*}

\begin{figure}[t]
    \centering
    \includegraphics[width=\columnwidth]{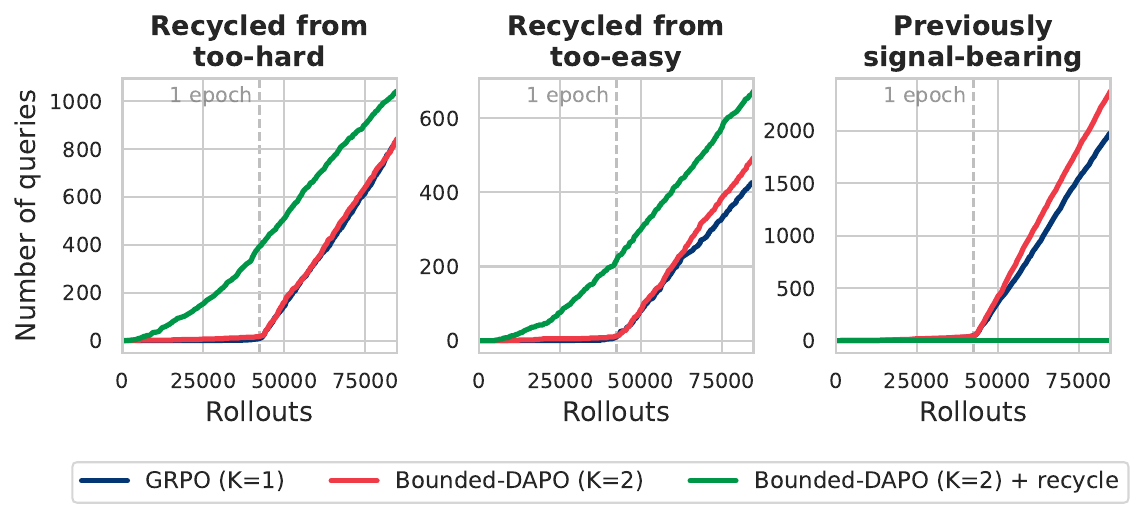}
    \vspace{-0.15cm}
    \caption{Cumulative queries seen during training on Qwen3-1.7B, decomposed by initial state: recycled from too-hard (left), recycled from too-easy (middle), and revisits to previously signal-bearing queries (right).}
    \vspace{-0.15cm}
    \label{fig:unique_queries}
\end{figure}

\begin{table}[t!]
\centering
\small
\setlength{\tabcolsep}{3.5pt}
\renewcommand{\arraystretch}{1.15}
\resizebox{\columnwidth}{!}{%
\begin{tabular}{lccccc}
\toprule
Model & $k$ & Recycle & GAIA & HLE & WW \\
\midrule
\rowcolor{gray!15}
\multicolumn{6}{l}{\textbf{Prior work}} \\
BehaviorPrime-1.7B  & -- & --        & 21.4 & 7.8 & \textbf{37.2}  \\
SmartSearch-3B    & -- & --        & 19.4 & 9.8 & 26.8 \\
ASearcher-Web-7B       & -- & --        & 22.3 & 7.0 & 21.5  \\
\midrule
\rowcolor{gray!15}
\multicolumn{6}{l}{\textbf{Ours} (Qwen3 + SFT + B-DAPO w/ recycling)} \\
\quad 1.7B & 2 & \ding{51} & 21.4 & 8.0 & 35.4  \\
\quad 4B   & 2 & \ding{51} & \textbf{24.8} & \textbf{10.4} & 36.3  \\
\bottomrule
\end{tabular}%
}
\vspace{-0.15cm}
\caption{Pass@1 on GAIA, HLE, and WebWalker.}
\vspace{-0.15cm}
\label{tab:main-results-small}
\end{table}

\paragraph{Agentic Search Results.}
Table~\ref{tab:main-results} reports Pass@1 results on the seven MHQA benchmarks. At the 1.7B parameter scale, our full pipeline achieves 66.0 average, matching or exceeding prior systems trained on benchmark-derived supervision. Our setup uses only synthetic queries, and the resulting agents transfer zero-shot along two axes: from synthetic training queries to different benchmark distributions, and from the sandboxed DeepResearchGym retrieval environment, used during training, to live SERPER search at inference time. At a matched rollout budget, recycling improves over both two-epoch GRPO and Bounded-DAPO. Scaling the same pipeline from Qwen3-1.7B to Qwen3-4B shows consistent results, improving performance to 71.3 average across the datasets.

\paragraph{Recycling Mechanism.}
Multi-epoch GRPO and Bounded-DAPO follow a fixed revisiting schedule: in a second epoch every query is sampled again, regardless of whether it produced signal in the first pass. Recycling replaces this with a status-dependent allocation, concentrating subsequent rollouts on queries that previously produced zero-variance groups. Figure~\ref{fig:unique_queries} decomposes the unique queries each method extracts signal from, by their initial state. The inflection at one epoch separates the two regimes. Without recycling, the second epoch is dominated by revisits, i.e. both GRPO and Bounded-DAPO accumulate rollouts on previously consumed queries. Recycling never revisits (rightmost panel, flat at zero) and instead recovers additional too-hard and too-easy queries under the same total rollouts.

\paragraph{Long-horizon Reasoning.}

Besides multi-hop question answering, Table~\ref{tab:main-results-small} shows similar trends on harder web-navigation and long-horizon reasoning benchmarks. Evaluating our recycle-based 1.7B and 4B models, performance remains competitive despite training exclusively on English synthetic data and evaluating on open-web environments. The main exception is WebWalker, whose multilingual setting differs from our English-only training distribution.

\subsection{Ablations and Analysis}
\label{sec:analysis}

This section presents a detailed analysis of the obtained results, with additional perspectives being presented in Appendix~\ref{app:1} and~\ref{app:2}.

\begin{figure}[t]
    \centering
    \includegraphics[width=\columnwidth]{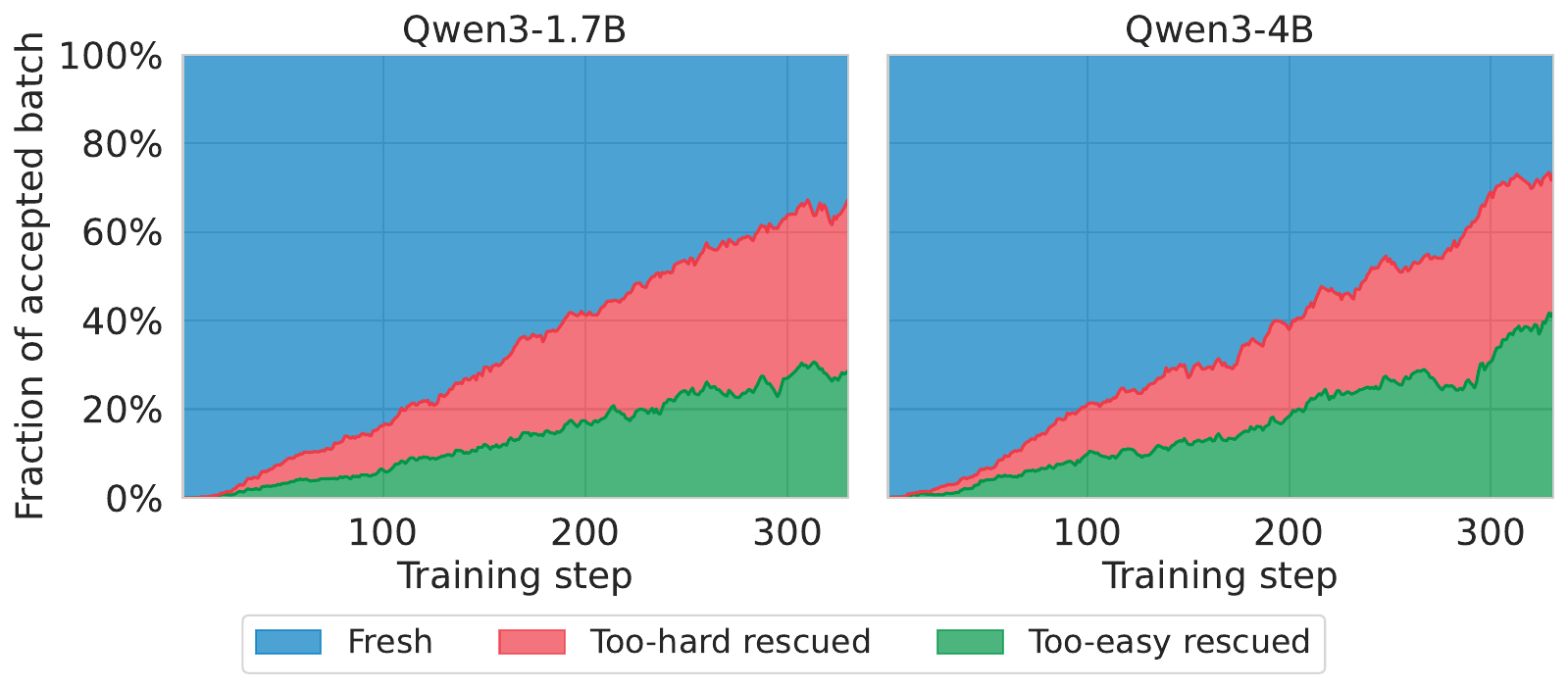}
    \caption{Composition of the effective batch (i.e. signal-bearing, $S_t$) over
    training on Qwen3 1.7B and 4B. \emph{Fresh} queries are drawn from the pool
    for the first time.}
    \label{fig:effective_composition}
    \vspace{-0.4cm}
\end{figure}

\paragraph{Recycled Queries Dominate the Effective Batch.}
Figure~\ref{fig:effective_composition} shows the composition of the effective training batch during recycling-based training. Early in training, most signal-bearing groups come from freshly sampled queries. As training progresses, recycled queries increasingly dominate the effective batch. By the final steps, rescued too-hard and too-easy queries together account for three quarters of accepted groups. Both rescue populations grow throughout training and neither saturates within our horizon, suggesting that informative queries continue to emerge from previously zero-variance regions. Too-hard and too-easy rescues appear at comparable scale, indicating that recycling captures delayed capability acquisition, and re-entry from previously mastered behavior.

\begin{figure}[t]
    \centering
    \includegraphics[width=\columnwidth]{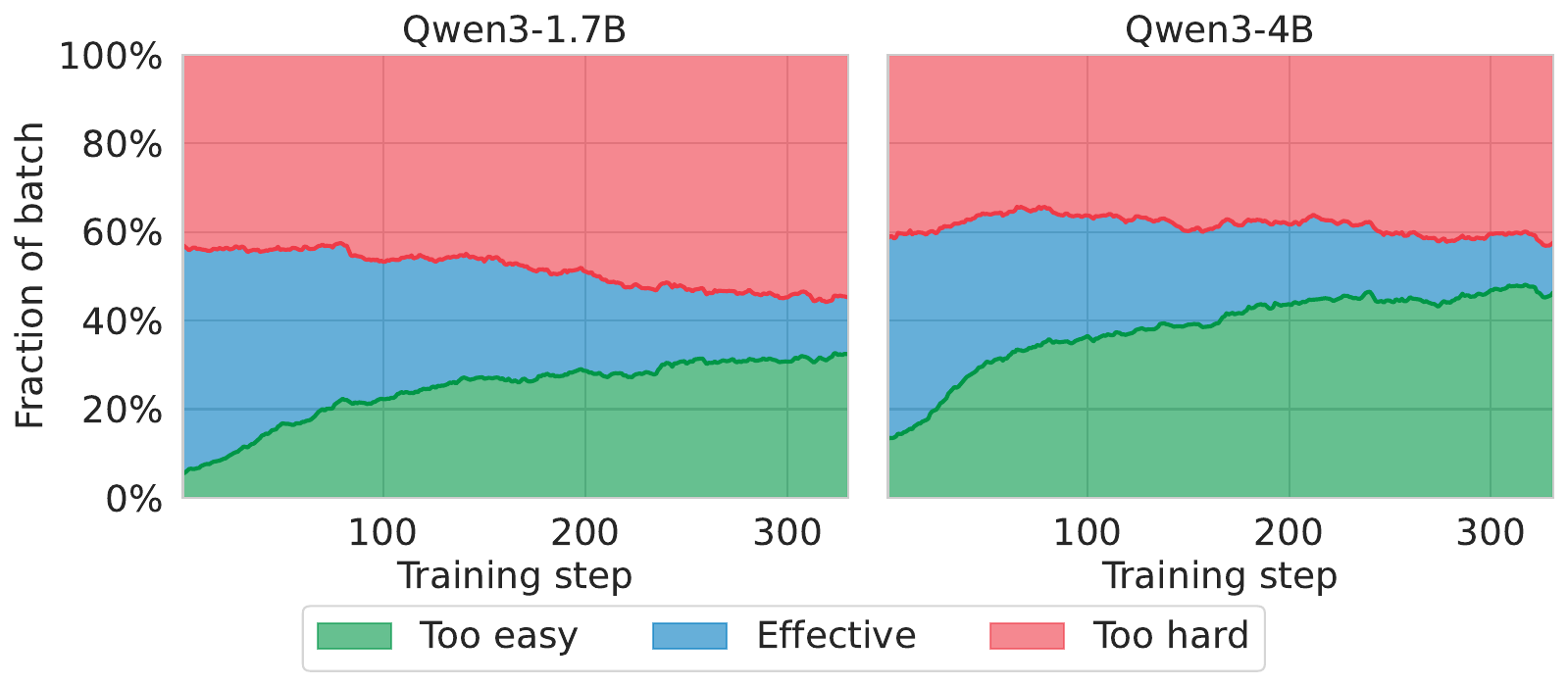}
    \vspace{-0.15cm}
    \caption{Composition of the sampled batch (i.e. all candidates, $C_t$) under recycling for Qwen3-1.7B (left) and Qwen3-4B (right).}
    \vspace{-0.15cm}
    \label{fig:pool_composition}
\end{figure}

\paragraph{Pool Difficulty Evolves with Model Capacity.}
Figure~\ref{fig:pool_composition} tracks the composition of the sampled pool under recycling throughout training. For the 1.7B model, the too-easy fraction increases steadily to approximately $35\%$, while the effective fraction gradually declines, and the too-hard region remains dominant throughout training. This suggests that the smaller model converts part of the pool into mastered queries, but leaves a large fraction of queries persistently beyond its capability.

In contrast, the 4B model exhibits a different dynamic, with a relativelly steady too-hard fraction, and a too-easy one growing to nearly $50\%$. These trends are consistent with a stronger policy progressively converting initially unsolved queries into solvable and eventually mastered ones, producing a clear migration towards the too-easy regions of the pool. This suggests that zero-variance difficulty is policy-relative. Under an otherwise identical training setup, the larger model converts a substantial fraction of previously unsolved queries into effective or mastered ones. This further indicates that the too-hard pool reflects capability limitations rather than a synthetic sampling noise floor.

\begin{figure}[t]
    \centering
    \includegraphics[width=\columnwidth]{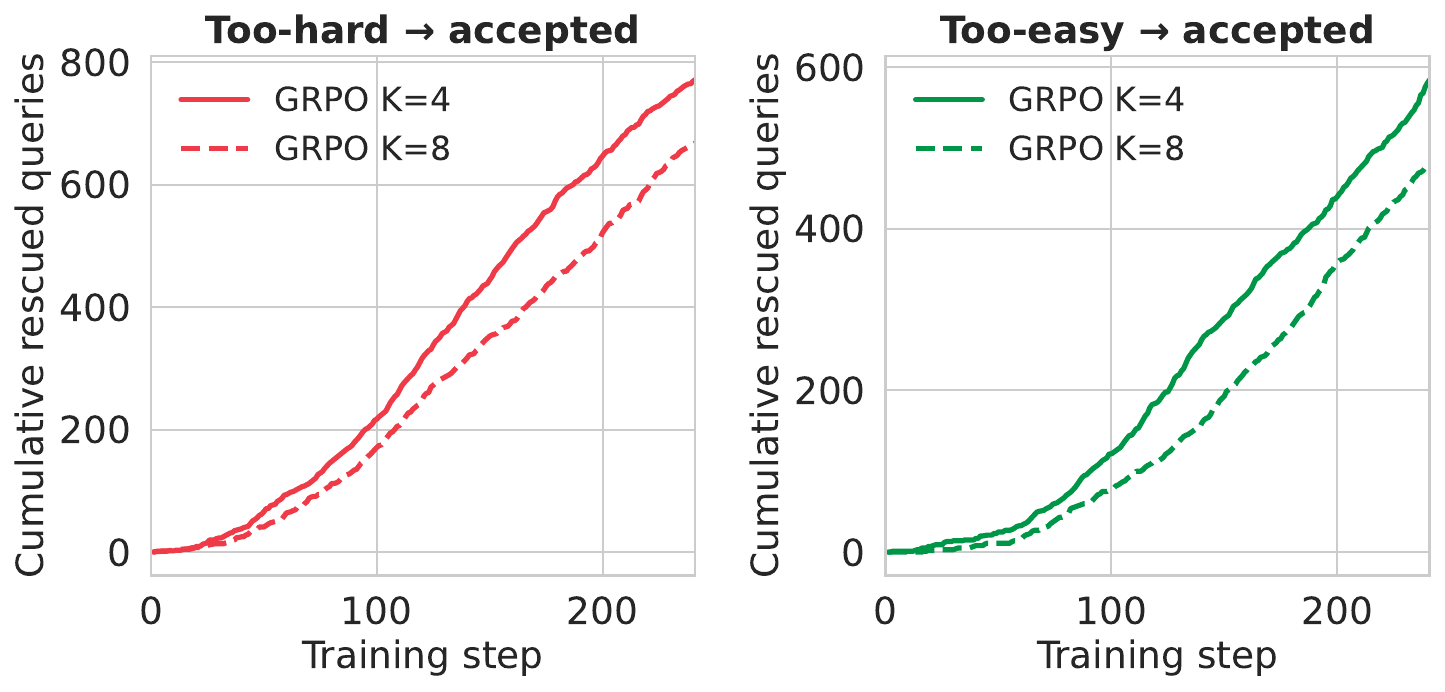}
    \vspace{-0.15cm}
    \caption{Cumulative recycled queries on Qwen3-1.7B at GRPO group size $K{=}4$ and
    $K{=}8$.}
    \vspace{-0.15cm}
    \label{fig:rescue_cumsum}
\end{figure}

% \paragraph{Group size as a difficulty estimator.}

% The per-group rollout count $K$ controls both rollout cost and the granularity of zero-variance classification. For a query $q$ with policy success probability $p_\pi(q)$, the probability of observing a zero-variance group is $p_\pi(q)^K + (1 - p_\pi(q))^K$. Larger $K$ therefore concentrates the zero-variance pool more tightly around the extreme tails of the difficulty distribution, while smaller $K$ admits more intermediate-difficulty queries that happened to produce all-same outcomes under finite sampling.

% Figure~\ref{fig:rescue_cumsum} reflects this effect empirically. Compared to $K{=}8$, the noisier $K{=}4$ setting produces approximatelly $17\%$ more too-hard rescues and $23\%$ more too-easy rescues, consistent with more queries being temporarily assigned to zero-variance regions due to sampling noise and later re-entering the effective set after resampling. Despite this noisier classification, final average accuracy over the 7 MHQA datasets at $K{=}4$ matches $K{=}8$ at half the rollout cost, making $K{=}4$ the more favorable compute trade-off in our setting.

\paragraph{Group Size as a Difficulty Estimator.}
The per-group rollout count $K$ controls both rollout cost and the granularity of zero-variance classification. From Equation ~\ref{eq:pzv}, larger $K$ concentrates the zero-variance pool around the extreme tails of the difficulty distribution, while smaller $K$ admits more intermediate-difficulty queries that happen to produce all-same outcomes under finite sampling.

Figure~\ref{fig:rescue_cumsum} reflects this effect empirically. Compared to $K{=}8$, the noisier $K{=}4$ setting produces approximately $17\%$ more too-hard rescues and $23\%$ more too-easy rescues, consistent with more queries being temporarily assigned to zero-variance regions due to sampling noise, and later re-entering the effective set after resampling. Despite this noisier classification, the final averaged accuracy over the 7 MHQA datasets at $K{=}4$ matches $K{=}8$ at half the rollout cost, making $K{=}4$ the more favorable compute trade-off in our setting.   
\vspace{-0.15cm}
\section{Conclusions and Future Work}
\vspace{-0.15cm}

In this paper, we studied query recycling as a train-time modification for LLM search agents that returns zero-variance rollout groups to the active pool for resampling at later training steps. On a 10k pool of synthetic queries, a 1.7B agent trained with recycling reaches 66.0 average Pass@1 across seven multi-hop QA benchmarks, matching or surpassing 7B systems trained on benchmark-derived supervision. Scaling to models with 4B parameters further improves performance to 71.3.

Our analysis shows that the zero-variance status is policy-relative across training steps and model scales. The 4B model converts close to half of the pool into too-easy queries by the end of training, while the 1.7B model leaves the too-hard region dominant throughout. Too-easy queries account for 40\% of all rescues, so recoverable signal arrives from policy drift alongside policy improvement. Methods that intervene only on all-failed groups would miss this contribution.

Recycling is also compatible with denser supervision signals and with advantage shaping over zero-variance groups, and we plan to study these combinations. A fixed pool also constrains the training horizon, and, as the policy improves, the signal-bearing fraction of the pool shrinks. Generating new queries online, matched to the policy's current capability, would relax this constraint. Yet another direction for future work concerns extending recycling to deep research systems that perform long-form report generation, where reward design becomes the harder problem.   
\section*{Limitations}

Our evaluation follows prior work in using an LLM judge for answer correctness. This enables direct comparisons and admits surface variation that strict exact-match would penalize, but it inherits reproducibility issues associated with the procedure. The usage of the SERPER live search engine for inference results also incurs reproducibility issues, although this is standard practice in the area.

Most of our evaluation datasets are English-only, and our synthetic query pool is likewise grounded in the English subset of ClueWeb22. Multilingual settings are thus out of distribution for our pipeline.

We study the Qwen3 model family at 1.7B and 4B parameters, and the recycling dynamics we report are observed at this scale. The balance between too-easy and too-hard rescues may shift at larger scales or with other model families, although tailoring the initial state of the synthetic pool to model capacity is a natural way to compensate.

Finally, our RL experiments leverage outcome-only rewards. Recycling is in principle orthogonal to denser supervision signals such as process rewards, but we do not test this combination.

\section*{Ethical Considerations}

All the evaluation benchmarks and base LLMs used in our experiments are publicly available for research use. We will open-source the code that allows for reproduction of the results, as well as model checkpoints and synthetic data.

By using large pre-trained language models, we acknowledge the risks associated with the presence of inherent biases embedded within the models, which may inadvertently perpetuate or amplify societal biases present in the training data.

\bibliography{custom.bib}

\appendix
\section{Training Dynamics Across the Different Synthetic Query Sources}
\label{app:1}

\begin{figure*}[t!]
    \centering
    \includegraphics[width=\linewidth]{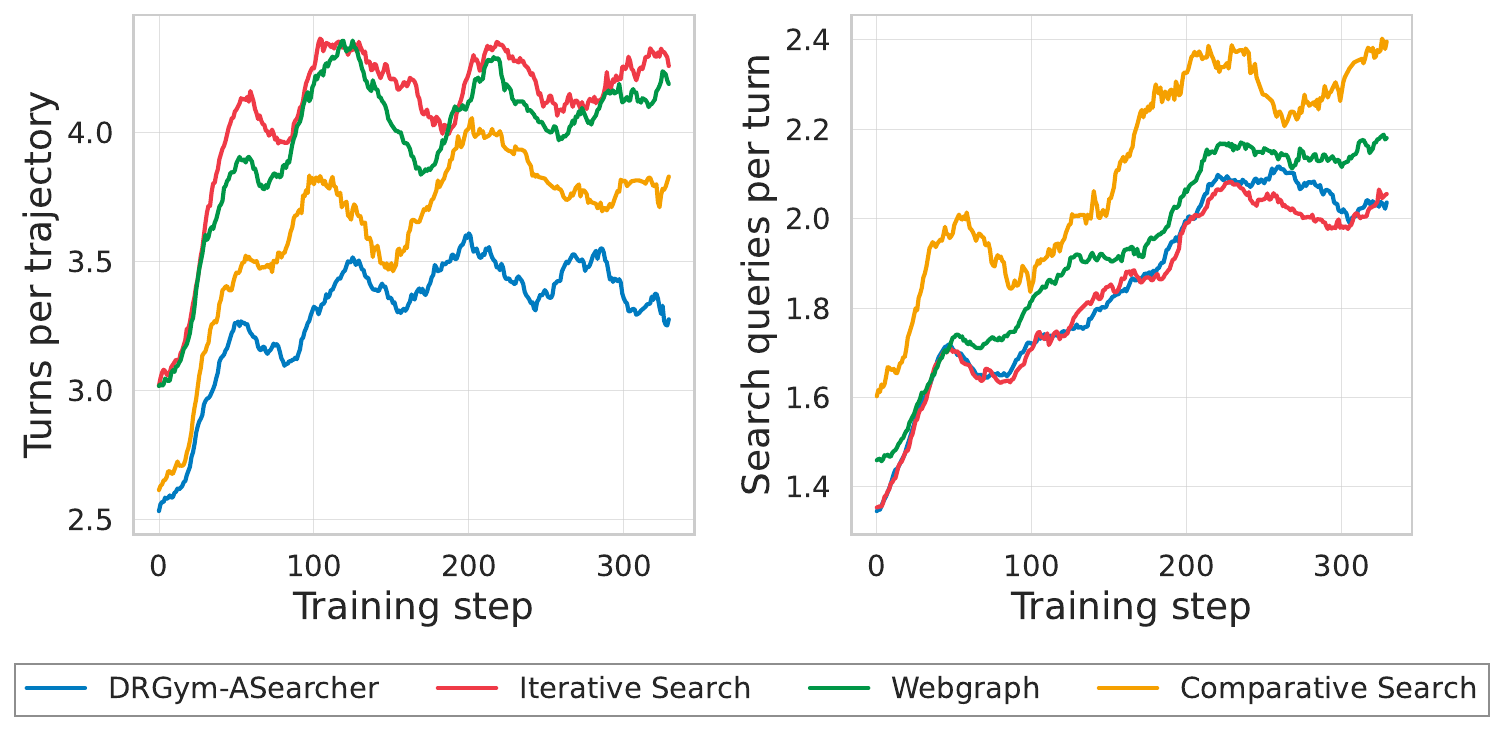}
    \caption{Average trajectory statistics during Qwen3-1.7B RL training under recycling, grouped by synthetic query source. Left: average number of turns per trajectory. Right: average number of search queries issued per turn. DRGym-ASearcher queries require consistently shorter trajectories, while comparative queries exhibit higher query parallelism, reflected in fewer turns but more search calls per turn.}
    \label{fig:search_Evo}
\end{figure*}

\begin{figure*}[t!]
    \centering
    \includegraphics[width=\linewidth]{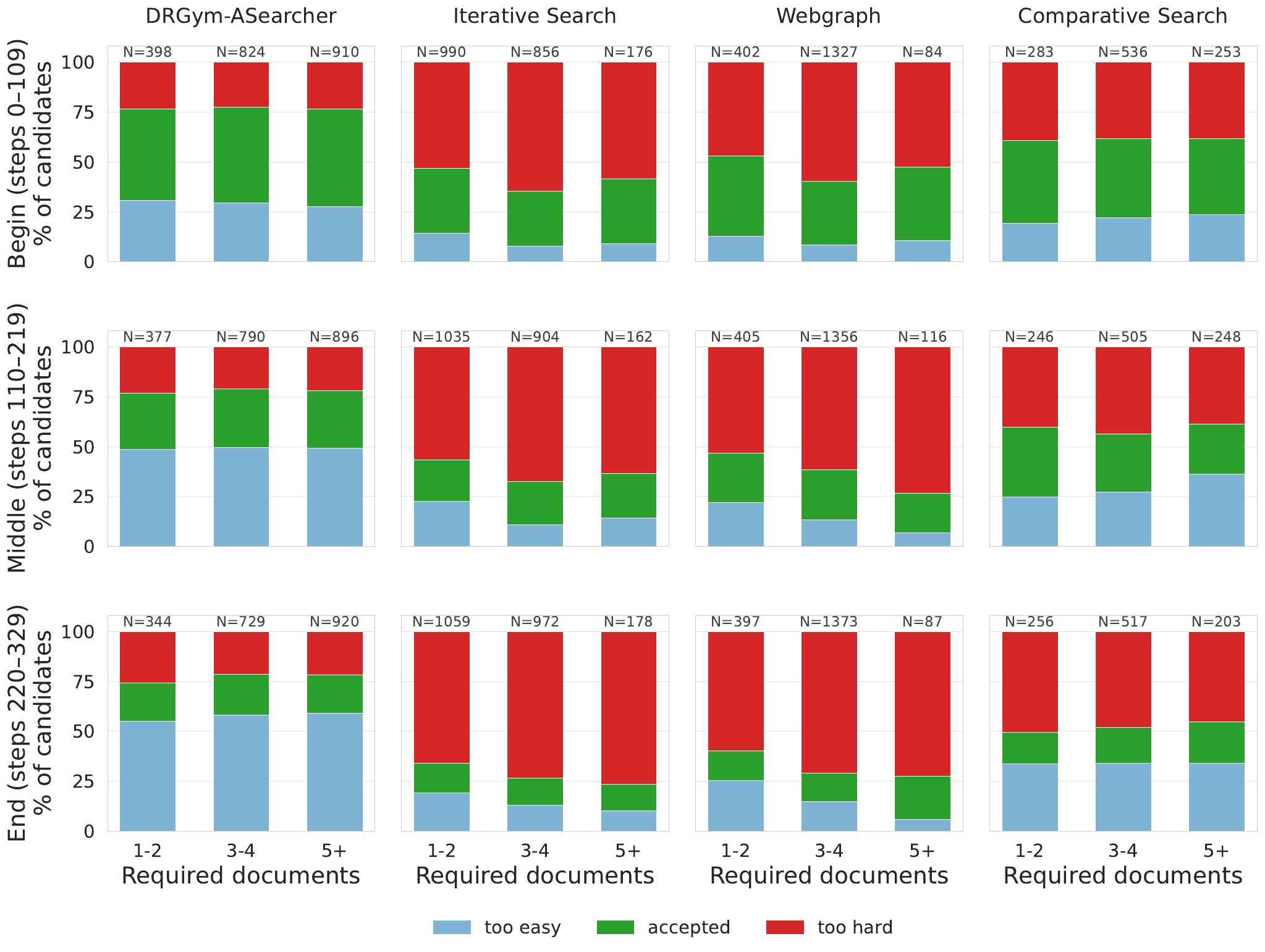}
    \caption{Composition of sampled candidate queries throughout Qwen3-1.7B recycling-based training, stratified by generation source and required document count. Each bar reports the fraction of sampled candidates classified as too-easy, accepted, or too-hard within a training stage (early, middle, or end). Values above bars indicate the number of sampled queries in each subgroup.}
    \label{fig:per_source_signal}
\end{figure*}

Figures~\ref{fig:search_Evo} and~\ref{fig:per_source_signal} decompose the training dynamics of the Qwen3-1.7B recycle run according to the query generation source and required document count, showing that the generation strategies induce distinct search behaviors. 

For instance, in Figure~\ref{fig:search_Evo}, DRGym-ASearcher queries require fewer turns per trajectory than the other sources, consistent with their higher too-easy fraction as shown in Figure~\ref{fig:per_source_signal}, especially early in training. Iterative-search and webgraph queries induce longer trajectories, reflecting their sequential multi-hop construction.

Comparative-search queries follow a different profile in
Figure~\ref{fig:search_Evo}. The agent solves them with fewer turns while issuing more search queries per turn, despite the absence of explicit supervision encouraging parallel retrieval. Comparative questions aggregate independently retrievable entities, and the policy dispatches these retrievals in parallel through the search tool interface.

Source-level difficulty in Figure~\ref{fig:per_source_signal} is already separated at the start of training. DRGym-ASearcher contains a large too-easy fraction from the first stage, while iterative-search and webgraph queries concentrate in the too-hard region, particularly at higher document counts.

As training progresses, the sampled pool shifts toward too-hard queries. However, this shift reflects pool composition rather than policy degradation. Recycling removes accepted queries from the active pool while leaving zero-variance queries eligible for resampling, and thus the remaining candidates concentrate on queries that the current policy does not yet solve consistently.

The aforementioned effect is most visible in the middle two columns of Figure~\ref{fig:per_source_signal}, where the accepted fraction of iterative-search and webgraph queries decreases while their too-hard fraction expands. Their longer sequential reasoning chains keep them near the frontier of the model's capabilities, and they may become signal-bearing once the policy improves enough to solve them.

\section{Soft Recycling}
\label{app:2}
Section~\ref{sec:rl} instantiates query recycling as a binary weight rule, i.e., consumed signal-bearing queries are removed from the pool ($w_q^{(t+1)} = 0$), and zero-variance queries are returned at full eligibility ($w_q^{(t+1)} = 1$). We now describe one relaxation, \emph{soft recycling}, which downweights consumed queries instead of removing them:
\begin{equation*}
w_q^{(t+1)} = \begin{cases}
\alpha & \text{if } q \in S_t, \\
1 & \text{otherwise},
\end{cases}
\end{equation*}
where $\alpha \in [0, 1]$ is a hyperparameter and $S_t$ is the set of signal-bearing queries selected into the gradient update batch at step $t$. Queries not drawn at step $t$ keep their previous weight. The binary rule of Section~\ref{sec:rl} corresponds to $\alpha = 0$.
\paragraph{Consumed queries.} Signal-bearing queries selected into the gradient update batch keep weight $\alpha$ instead of being removed. With $\alpha > 0$ they continue to compete for sampling mass at later steps.

\begin{table*}[h]
\small
\setlength{\tabcolsep}{4pt}
\renewcommand{\arraystretch}{1.15}
\begin{tabular}{lllcccccccccc}
\toprule
Model & Size & $k$ & Rollouts & 2Wiki & Bamb & HQA & MSQ & NQ & PopQA & TriQA & Avg \\
\midrule
\rowcolor{gray!15}
\multicolumn{12}{l}{\textbf{Ours} (full synthetic training over ClueWeb22 and evaluation using SERPER)} \\
\addlinespace[3pt]
\cdashline{1-12}
\addlinespace[3pt]
Qwen3                            & 1.7B & -- & --    & 37.4 & 44.4 & 44.1 & 16.3 & 54.5 & 41.8 & 68.6 & 43.9 $\pm$ 0.40 \\
\addlinespace[3pt]
\cdashline{1-12}
\addlinespace[3pt]
\quad + SFT                      & 1.7B & -- & --    & 54.8 & 47.2 & 54.4 & 24.6 & 67.6 & 52.0 & 86.0 & 55.2 $\pm$ 0.33 \\
\quad + GRPO                     & 1.7B & 1  & 84.5K & \underline{70.9} & 68.8 & 66.6 & \underline{36.9} & \underline{69.9} & \underline{54.4} & 86.9 & 64.9 $\pm$ 0.21 \\
\quad + B-DAPO                   & 1.7B & 2  & 84.5K & 69.5 & 67.2 & 66.9 & \underline{36.9} & 69.5 & 53.5 & 85.2 & 64.1 $\pm$ 0.20 \\
\quad + B-DAPO w/ recycling      & 1.7B & 2  & 84.5K & \textbf{71.7} & \textbf{72.4} & \textbf{67.7} & \textbf{38.0} & \textbf{70.1} & 54.2 & \underline{87.8} & \textbf{66.0} $\pm$ 0.17 \\
\quad + B-DAPO w/ soft recycling & 1.7B & 2  & 84.5K & 69.6 & \underline{71.1} & \underline{67.3} & 36.6 & 69.4 & \textbf{54.7} & \textbf{88.1} & \underline{65.2} $\pm$ 0.20 \\
\bottomrule
\end{tabular}
\caption{Pass@1 on seven multi-hop QA benchmarks for the 1.7B model. All RL configurations start from the SFT checkpoint and use a matched rollout budget. Soft recycling uses $\alpha = 0.1$. Averages over inference runs are reported as mean $\pm$ standard error.}
\label{tab:soft-recycling}
\end{table*}

\paragraph{Zero-variance queries.} Zero-variance queries are returned at full eligibility ($w_q^{(t+1)} = 1$), as in the binary rule.
We report a run with $\alpha = 0.1$, so that consumed queries remain sampleable at one tenth the weight of fresh queries. All other training settings match the Qwen3-1.7B configuration of Section~\ref{sec:exp_method}. Table~\ref{tab:soft-recycling} reports the result. Soft recycling reaches 65.2 average Pass@1, above both non-recycling baselines in Table~\ref{tab:main-results} but below the 66.0 of the binary rule. Down-weighting consumed queries rather than removing them therefore recovers part of the recycling gain, though it trails the binary rule at $\alpha = 0.1$. We did not sweep $\alpha$, so we do not rule out gains at other values.

\end{document}